\documentclass[superscriptaddress, twocolumn, prl, longbibliography, nofootinbib]{revtex4-1}
\usepackage{amsmath}
\usepackage{amsbsy}
\usepackage{amssymb}
\usepackage{graphicx}
\usepackage{color}
\usepackage{subfigure}
\usepackage{physics}
\usepackage{soul}
\usepackage{color}
\usepackage{bm}
\usepackage{array}
\usepackage{multirow}
\usepackage{lipsum}
\usepackage[normalem]{ulem}
\usepackage{verbatim}
\usepackage{natbib}
\usepackage{nccmath}
\definecolor{linkcolor}{rgb}{0,0,0.6} 
\usepackage[pdftex,colorlinks=true,
	pdfstartview = FitV,
	linkcolor    = linkcolor,
	citecolor    = linkcolor,
	urlcolor     = linkcolor,	
	hyperindex   = true,
	hyperfigures = false]{hyperref}

\newcommand{\bP}{\text{\bf P}}
\newcommand{\T}{T}
\newcommand{\U}{U}

\newcommand\numberthis{\addtocounter{equation}{1}\tag{\theequation}}

\begin{document} 

\title{Mean-field theory for the structure of strongly interacting active liquids}

\author{Laura Tociu}
\author{Gregory Rassolov}
\affiliation{James Franck Institute, University of Chicago, Chicago, IL 60637}
\affiliation{Department of Chemistry, University of Chicago, Chicago, IL 60637}

\author{\'Etienne Fodor}
\affiliation{Department of Physics and Materials Science, University of Luxembourg, L-1511 Luxembourg}

\author{Suriyanarayanan Vaikuntanathan}
\affiliation{James Franck Institute, University of Chicago, Chicago, IL 60637}
\affiliation{Department of Chemistry, University of Chicago, Chicago, IL 60637}
\begin{abstract}
Active systems, which are driven out  of equilibrium by local non-conservative forces, exhibit unique behaviors and structures with potential utility for the design of novel materials. An important and difficult challenge along the path towards this goal is to precisely predict how the structure of active systems is modified as their driving forces push them out of equilibrium. Here, we use tools from liquid-state theories to approach this challenge for a classic minimal active matter model. First, we construct a nonequilibrium mean-field framework which can predict the structure of systems of weakly interacting particles. Second, motivated by equilibrium solvation theories, we modify this theory to extend it with surprisingly high accuracy to systems of strongly interacting particles, distinguishing it from most existing similarly tractable approaches. Our results provide insight into spatial organization in strongly interacting out-of-equilibrium systems.
\end{abstract}

\maketitle
{A}ctive matter is a class of nonequilibrium systems in which every component consumes energy to produce an autonomous motion~\cite{Marchetti2013, Bechinger2016, Marchetti2018}. Examples of active systems span many length- and time-scales, from bacterial swarms~\cite{Elgeti2015} and assemblies of self-propelled colloids~\cite{Palacci2013}, to animal groups~\cite{Cavagna2014} and human crowds~\cite{Bartolo2019}. The energy fluxes stemming from individual self-propulsion lead to complex collective behaviors without any equilibrium equivalent, such as collective directed motion~\cite{Sood2014} and phase separation despite purely repulsive interactions~\cite{Palacci2013}. The possibility of exploiting such behaviors to design materials with innovative functions has motivated much research~\cite{Geiss2009}, with the goal of reliably predicting and controlling the features of active systems.

Minimal models have been proposed to capture active dynamics of particles with aligning interactions and of self-propelled isotropic particles, which yield collective motion~\cite{Chate2020} and motility-induced phase separation~\cite{Cates2015} respectively. Based on these models, the challenge is to establish a nonequilibrium framework, by analogy with equilibrium statistical thermodynamics, which connects microscopic details and emergent physics. Progress has been made in this direction by characterizing protocol-based observables, such as pressure~\cite{Brady2014, Solon2015}, surface tension~\cite{Speck2015, Zakine2020}, and chemical potential~\cite{Guioth2019}.

Despite recent advancements, understanding how to quantitatively control the dynamics and structure of many-body active systems by appropriately tuning external parameters remains largely an open challenge~\cite{Gompper_2020}. A large part of the theoretical approaches used to predict the structure of active fluids generally rely on either equilibrium mappings~\cite{Szamel2015, Rein2016, Wittmann2017b, Szamel2019} or weak-interaction approximations~\cite{Suri2019, Suri2020}, thus limiting their applicability.

In this work, we use tools from liquid-state theories to take on this challenge. We construct a novel mean-field theory whose applicability and ease of implementation surpasses existing approaches, and which quantitatively predicts the static two-point density correlations in a minimal isotropic active matter system both near and far from equilibrium. Our results illustrate how the structure of a nonequilibrium many-body system can be controlled by tuning their driving forces. In later work, we develop expressions connecting these two-point density correlations to energy dissipation for the same active matter system and in more complex anisotropic systems, and demonstrate how artificial intelligence can potentially be harnessed to tune the structure of such systems.

The paper is organized as follows. First, we describe the model of an active liquid that we analyze for the majority of the manuscript, which is an assembly of self-propelled particles. Second, we outline calculations that accurately solves for the structure of our active liquid both near and far from equilibrium when particles are {\it weakly interacting}. Third, motivated by equilibrium solvation theories \textendash~these have shown how the equilibrium structure of liquids can be resolved by separately considering the rapidly varying and slowing components of the interaction~\cite{lum1999} \textendash~and by measurements from simulation of the relaxation of strongly interacting particles at equilibrium in response to perturbations, we develop a novel non-equilibrium mean-field theory for strongly interacting particles~\eqref{eq:final_expression_gr_c}, effectively using the direct correlation function of the system at equilibrium to account for higher-order interactions. This is the main result of this paper. Unlike many other reasonably accurate representations of active dynamics~\cite{Szamel2015, Rein2016, Wittmann2017, Szamel2019}, our final results do not rely on any equilibrium approximation, thus allowing all nonequilibrium features to be retained.


\section{Results}

\subsection{Details of model active matter system}

We consider a popular model of active matter consisting of $N$ interacting self-propelled particles, often referred to as Active Ornstein Uhlenbeck Particles~\cite{Szamel2014, Maggi2015, Nardini2016}, with two-dimensional overdamped dynamics:
\begin{equation}\label{eq:EOM}
	\dot{\bf r}_i = -\frac{1}{\gamma}\nabla_i \sum_{j\neq i} U({\bf r}_i-{\bf r}_j) + \frac{{\bf f}_i}{\gamma} + {\boldsymbol\xi}_i ,
\end{equation}
where $U$ is the pair-wise potential and $\gamma$ is a friction coefficient. The terms $\{{\boldsymbol\xi}_i,{\bf f}_i\}$ embody, respectively, the thermal noise and the self-propulsion velocity. They have Gaussian statistics with zero mean and uncorrelated variances, given by:
\begin{align}\label{eq:EOM_noise}
    & \langle\xi_{i\alpha}(t)\xi_{j\beta}(0)\rangle = \frac{2T}{\gamma} \delta_{ij}\delta_{\alpha\beta}\delta(t) \nonumber , \\
    & \langle f_{i\alpha}(t) f_{j\beta}(0) \rangle = \frac{\gamma T_{\rm A}}{\tau} \delta_{ij} \delta_{\alpha\beta} e^{-|t|/\tau} ,
\end{align}
where $\tau$ is the persistence time. For a vanishingly small persistence ($\tau \rightarrow 0$), the system reduces to a set of passive Brownian particles at temperature $T+T_{\rm A}$. At sufficiently high persistence and $\tau$, the system undergoes phase separation even with a purely repulsive interparticle potential $U$~\cite{Nardini2016}. All details pertaining to the simulations, run in two dimensions for all of what follows, can be found in Materials and Methods.

\subsection{Density field for a weakly interacting tracer particle}

We start by considering the effective dynamics of an active tracer embedded in a bath consisting of the other particles. To analytically derive the statistics of the tracer displacement, our strategy, inspired by recent works~\cite{Demery2011, Demery2014}, is to rely on a mean-field approach by first considering that interactions between the tracer and the bath are weak. This leads us to scale the interaction strength by a dimensionless factor $\varepsilon$, which can be regarded as a small parameter for perturbative expansion. The equation of motion of the tracer position ${\bf r}_0(t)$ then reads
\begin{equation}\label{eq:dyn_pert}
    \dot{\bf r}_0 = {\bf f}_0 - \varepsilon \int \nabla_0 \U({\bf r}_0  - {\bf r}') \rho ({\bf r}', t)  d {\bf r}' + {\bm \xi}_0 ,
\end{equation}
where the bath is described in terms of the density field $\rho ( {\bf r} , t) = \sum_{i=1}^N \delta({\bf r} - {\bf r}_i(t))$ with $N$ the number of bath particles. Note that we set $\gamma = 1$ here and in all subsequent equations. The dynamics of the density field $\rho({\bf r},t)$ can be obtained following the procedure in~\cite{Dean_1996}:
\begin{align*}\label{eq:field_EOM}
    &\dfrac{\partial \rho ({\bf r}, t)}{ \partial t} = \T \nabla^2 \rho({\bf r}, t) + \nabla \cdot \big[ \sqrt{2 \rho \T} {\bm \Lambda}({\bf r}, t) -\bP({\bf r}, t) \big] 
    \\
    &\quad + \nabla \cdot \left( \rho \nabla \left[ \int  \U({\bf r}  - {\bf r}') \rho({\bf r}', t)  d {\bf r}' + \varepsilon\U({\bf r} - {\bf r}_0)  \right] \right) , \numberthis
\end{align*}
where $\bP$ denotes here the polarization field $ \bP ({\bf r}, t) = \sum_i {\bf f}_i(t) \delta({\bf r} - {\bf r}_i(t))$. The term $\boldsymbol\Lambda$ is a Gaussian white noise with zero mean and unit variance ($\langle \Lambda_{\alpha} ({\bf r}, t) \Lambda_{\beta} ({\bf r}', t') \rangle = \delta_{\alpha \beta} \delta({\bf r} - {\bf r}') \delta(t - t')$). In principle, the dynamics~(\ref{eq:dyn_pert}-\ref{eq:field_EOM}) can be solved recursively to obtain the statistics of the density field $\rho({\bf r} , t)$ and of the tracer position ${\bf r}_0$. Some of us already took this approach in~\cite{Suri2019, Suri2020} using a perturbation in the weak interaction limit. In what follows, we extend this approach to characterize the system beyond the regime of weak interactions.


\subsection{Mean-field theory for nonequilibrium structure of weakly interacting particles}

The structure of the system is determined by the two-point correlation of density $h$, defined by $\rho_0 h({\bf r} ) = (1/N) \sum_{i\neq j} \langle \delta({\bf r}-{\bf r}_i+{\bf r}_j) \rangle - \rho_0$, where $\rho_0$ denotes the overall average density. In the homogeneous state, where density correlations are evaluated by measuring the average number of particles away from {\it any} representative tracer, the Fourier transform $h({\bf k}) = \int {\rm e}^{i{\bf k}\cdot{\bf r}} h({\bf r})d{\bf r}$ can be written in terms of $\delta\rho=\rho-\rho_0$ as
\begin{equation}\label{eq:gk}
	h({\bf k}) = \frac{1}{\rho_0} \left \langle  e^{i{\bf k} \cdot {\bf r}_0 (t) } \delta\rho ({\bf k}, t) \right \rangle .
\end{equation}
Our nonequilibrium mean-field theory to solve for $h(\bf{k})$ is built as follows. We first linearize the dynamics~(\ref{eq:dyn_pert}-\ref{eq:field_EOM}) and obtain a solution for $\delta \rho$ in the Fourier domain. We next construct an expansion in the coupling parameter $\varepsilon$ to compute $h({\bf k})$ up to first order in $\varepsilon$. As mentioned previously, this form is valid only in the regime of weak interactions. In the following section, we go beyond this regime by drawing inspiration from equilibrium solvation theories~\cite{Chandler1993}.

We begin with Eq.~\eqref{eq:field_EOM} and do not consider the polarization term further. Our choice is justified in the low-activity limit and, beyond that, supported by the results in Ref.~\cite{Suri2020}. In that work, the formulas for efficiency and mobility were obtained by setting two-point polarization correlators to zero (see Eqs.~(8-9) in Appendix~A), and their results agree with data from simulations very closely even in systems with strong driving forces.

By ignoring polarization and by linearizing the dynamics of the density field $\rho$ around the overall density $\rho_0$, we arrive at closed-form equation of motion for $ \delta\rho = \rho - \rho_0$. This linear approximation holds when interparticle potentials are weak such that any local density fluctuation is small compared to $\rho_0$. The solution for $\delta \rho({\bf k}, t) = \int [ \rho({\bf r}, t) - \rho_0 ] e^{-i{\bf k} \cdot {\bf r}} d{\bf r}$ follows readily as
\begin{align*}\label{eq:red_field_EOM_k}
    \delta &\rho( {\bf k}, t) = \int_{-\infty}^t ds e^{-{\bf k}^2 G({\bf k}) (t-s)} \\
    & \times \left( -{\bf k}^2 \rho_0 \varepsilon \U({\bf k}) e^{-i{\bf k} \cdot {\bf r}_0(s)} + i{\bf k} \cdot \sqrt{2\rho_0 \T} \bm{\Lambda} ({\bf k}, s) \right), \numberthis
\end{align*}
where $G({\bf k}) =  T + \rho_0 U({\bf k})$, and $\boldsymbol\Lambda$ is a zero-mean Gaussian white noise with correlations
\begin{equation}\label{eq:field_noise_k}
    \langle \Lambda_{\alpha}({\bf k}, s)  \Lambda_{\beta}({\bf k}', s')  \rangle = (2\pi)^d \delta_{\alpha \beta}\delta(s-s')\delta({\bf k} + {\bf k}') ,
\end{equation}
where $d$ is the spatial dimension. Substituting~\eqref{eq:red_field_EOM_k} into~\eqref{eq:gk} yields
\begin{align}\label{eq:hk}
    \rho_0 &h({\bf k}) = \nonumber \bigg \langle e^{i{\bf k} \cdot {\bf r}_0(0) }  \int_{-\infty}^0 ds e^{{\bf k}^2 G({\bf k})s} \nonumber \\
    & \times\left[ - \rho_0 {\bf k}^2 \varepsilon U({\bf k}) e^{-i{\bf k} \cdot {\bf r}_0(s)} + i{\bf k}\cdot \sqrt{2\rho_0 \T}  {\bm \Lambda}({\bf k}, s) \right] \bigg \rangle .
\end{align}
Next, we solve for the tracer position ${\bf r}_0$ to yield an expression which depends instead on driving forces ${\bf f}_0$, thermal noise ${\bm \xi}_0$, and interparticle potential $\U({\bf k})$. From the tracer dynamics~\eqref{eq:dyn_pert}, we deduce
\begin{align}\label{eq:tracer}
    {\bf r}_0(0) &= \int_{-\infty}^0 [ {\bf f}_0(x) + {\bm \xi}_0(x)] dx  \nonumber \\
    & + \varepsilon \int \frac{d{\bf k}'}{(2\pi)^d} i{\bf k}' \U({\bf k}')\int_{-\infty}^0 ds' \,\delta \rho({\bf k}', s') e^{ i {\bf k}\cdot{\bf r}_0(s')} .
\end{align}
From this, after expanding with respect to the parameter $\varepsilon$, we derive
\begin{widetext}
\begin{equation}\label{eq:tracer_bis}
    e^{i{\bf k}\cdot{\bf r}_0(0)} = e^{i{\bf k}\cdot \int_{-\infty}^0 [ {\bf f}_0(x) + {\bm \xi}_0(x)] dx} \bigg[ 1 - \varepsilon \int \frac{d{\bf k}'}{(2\pi)^d} {\bf k}\cdot{\bf k}' \U({\bf k}') \int_{-\infty}^0 ds'  \delta \rho({\bf k}', s') e^{i{\bf k}'\cdot \int_{-\infty}^{s'} [ {\bf f}_0(x) + {\bm \xi}_0(x)] dx} + {\cal O}(\varepsilon^2) \bigg] .
\end{equation}
Substituting in the expression for $\delta\rho$ given in Eq.~\eqref{eq:red_field_EOM_k}, we obtain
\begin{equation}\label{eq:tracer_ter}
\begin{aligned}
    e^{i{\bf k}\cdot{\bf r}_0(0)} = e^{i{\bf k}\cdot \int_{-\infty}^0 [ {\bf f}_0(x) + {\bm \xi}_0(x)] dx} \bigg[ 1 - &\varepsilon \int \frac{d{\bf k}'}{(2\pi)^d} ({\bf k}\cdot{\bf k}') \U({\bf k}') \int_{-\infty}^0 ds' e^{i{\bf k}'\cdot \int_{-\infty}^{s'} [ {\bf f}_0(x) + {\bm \xi}_0(x)] dx} \\
    &\times \int_{-\infty}^{s'} ds'' e^{-{\bf k}'^2 G({\bf k}') (s'-s'')} i {\bf k'} \cdot \sqrt{2T\rho_0} {\bm\Lambda}({\bf k}',s'') + {\cal O}(\varepsilon^2) \bigg] .
\end{aligned}
\end{equation}
Finally, substituting~\eqref{eq:tracer_ter} into~\eqref{eq:hk}, and expanding only to first order in $\varepsilon$, we derive
\begin{align}\label{eq:hk_bis}
    \rho_0 h({\bf k}) &= \bigg \langle -e^{i{\bf k}\cdot \int_{-\infty}^0 [ {\bf f}_0(x) + {\bm \xi}_0(x)] dx} \bigg[ \int_{-\infty}^0 ds e^{{\bf k}^2 G({\bf k})s} \rho_0 {\bf k}^2 \varepsilon U({\bf k}) e^{-i{\bf k}\cdot \int_{-\infty}^s [ {\bf f}_0(x) + {\bm \xi}_0(x)] dx} \bigg] \bigg \rangle \nonumber \\
    &\quad+ \bigg \langle e^{i{\bf k}\cdot \int_{-\infty}^0 [ {\bf f}_0(x) + {\bm \xi}_0(x)] dx} \bigg[ 1 - \varepsilon \int \frac{d{\bf k}'}{(2\pi)^d} ({\bf k}\cdot{\bf k}') \U({\bf k}') \int_{-\infty}^0 ds' e^{i{\bf k}'\cdot \int_{-\infty}^{s'} [ {\bf f}_0(x) + {\bm \xi}_0(x)] dx} \nonumber \\
    &\qquad\qquad \times\int_{-\infty}^{s'} ds'' e^{-{\bf k}'^2 G({\bf k}') (s'-s'')} i {\bf k'} \cdot \sqrt{2T\rho_0} {\bm\Lambda}({\bf k}',s'') \bigg] \bigg[ \int_{-\infty}^0 ds e^{{\bf k}^2 G({\bf k})s} i{\bf k}\cdot \sqrt{2\rho_0 \T}  {\bm \Lambda}({\bf k}, s) \bigg] \bigg \rangle.
\end{align}
We begin simplifying this expression by noting that $\langle {\bm \Lambda}({\bf k}, s) \rangle$ is zero, and making use of the fact that $\bm\Lambda$, ${\bm\xi}_0$, and ${\bf f}_0$ are independent, we eliminate the term that is order 0 in $\varepsilon$:
\begin{align}\label{eq:hk_ter}
    \rho_0 h({\bf k}) & = - {\bf k}^2 \varepsilon \rho_0 U({\bf k}) \int_{-\infty}^0 ds e^{{\bf k}^2 G({\bf k})s} \bigg \langle e^{i{\bf k}\cdot \int_{-\infty}^0 [ {\bf f}_0(x) + {\bm \xi}_0(x)] dx - i{\bf k}\cdot \int_{-\infty}^s [ {\bf f}_0(x) + {\bm \xi}_0(x)] dx} \bigg \rangle \nonumber \\
    & - 2 \rho_0 \T \int \frac{d{\bf k}'}{(2\pi)^d} ({\bf k}\cdot{\bf k}')^2 \varepsilon U({\bf k}') \int_{-\infty}^0 ds' e^{ -{\bf k}'^2 G({\bf k}') s'}  \Big\langle e^{i{\bf k}\cdot \int_{-\infty}^0 [{\bf f}_0(x) + {\bm \xi}_0(x) ]dx + i{\bf k}'\cdot \int_{-\infty}^{s'} [ {\bf f}_0(x) + {\bm \xi}_0(x)] dx} \Big\rangle \nonumber \\
    & \times \int_{-\infty}^{0} ds  \int_{-\infty}^{s'} ds'' e^{ {\bf k}^2 G({\bf k}) s + {\bf k}'^2 G({\bf k}') s''}  \langle \Lambda_\alpha({\bf k}, s) \Lambda_\alpha({\bf k}', s'')\rangle .
\end{align}
\end{widetext}
We further simplify this by observing that according to Wick's theorem, we can write for the white noise
\begin{equation}\label{eq:expxi}
    \left \langle e^{i{\bf k}\cdot \int_{s}^0 {\bm \xi}_0(x) dx} \right\rangle =e^{{\bf k}^2 Ts} .
\end{equation}
To treat the equivalent terms for the active forces, we start from the time correlations~\eqref{eq:EOM_noise} and derive the following:
\begin{equation}
    \left \langle \int_{s}^0 f_{0\alpha}(0) f_{0\alpha} (x) dx \right \rangle = T_{\rm A} (1 - e^{s/\tau}) ,
\end{equation}
\begin{equation}
\begin{aligned}\label{eq:expxi2}
    \bigg \langle \int_{s}^0 \int_{s}^0& f_{0\alpha} (x) f_{0\alpha} (x')  dx dx' \bigg\rangle \\
    &= \frac{T_{\rm A}}{\tau} \left( \int_{s}^0 \int_{x}^0 e^{-(x'-x)/\tau} dx' dx \right.
    \\
    &\quad +\left.\int_{s}^0 \int_{s}^x e^{-(x-x')/\tau} dx' dx \right) \\
    & = - 2 \big[ T_{\rm A} s + T_{\rm A} \tau ( 1- e^{s/\tau}) \big] \equiv - 2 R(s) .
\end{aligned}
\end{equation}
Again according to Wick's theorem, we can now write for the driving forces:
\begin{equation}\label{eq:expf}
    \left\langle e^{i{\bf k}\cdot \int_{s}^0 {\bf f}_0(x) dx} \right\rangle = e^{{\bf k}^2 R(s)} .
\end{equation}
In turn, this means that we can make the following simplification:
\begin{multline}
    \bigg \langle e^{i{\bf k}\cdot \int_{-\infty}^0 [ {\bf f}_0(x) + {\bm \xi}_0(x)] dx - i{\bf k}\cdot \int_{-\infty}^s [ {\bf f}_0(x) + {\bm \xi}_0(x)] dx} \bigg \rangle \\
    = e^{{\bf k}^2 \left(Ts + R(s)\right)} .
\end{multline}
After collapsing noise correlation functions and using Eqs.~(\ref{eq:field_noise_k}, \ref{eq:expxi}, \ref{eq:expf}) in this way to simplify~\eqref{eq:hk_ter}, we obtain that the form of the pair correlation function is
\begin{equation}\label{eq:final_expression_gr}
    h({\bf k}) = -  {\bf k}^2 \varepsilon U({\bf k})  \dfrac{ G({\bf k}) + T}{G({\bf k})} \int_{-\infty}^0 ds e^{{\bf k}^2 \left( (G({\bf k}) + T)s + R(s)\right)} .
\end{equation}
We set $\varepsilon$ to 1 to obtain an expression valid for systems with weak interparticle potentials and high densities.

\subsection{Mean-field theory for nonequilibrium structure of strongly interacting particles}

As mentioned previously, we go beyond the regime of weak interactions by drawing inspiration from equilibrium solvation theories~\cite{Chandler1993}. In this context, the density around a tracer particle interacting strongly with its neighbors is captured by considering the convolution between the density correlation and equilibrium direct correlation functions. The equilibrium direct correlation function can be readily obtained from the pair correlation function through the Ornstein-Zernike relation, and in the weak interaction limit, the direct correlation function is simply equal to the negative of the interparticle potential: $c_{\rm eq}({\bf r})=-U({\bf r})/T$~\cite{Hansen}. 
Hence, linear response in the weak interaction regime enforces that this convolution captures the same information as convoluting the density correlation function and interaction potential. Further, as has been demonstrated in the context of theories of the hydrophobic effect, the effect of any weaker perturbations can be handled by a mean field approach~\cite{lum1999} by perturbing around the equilibrium direct correlation function. In our context, intuition from these theories suggests the substitution of $U({\bf k})$ with $- T c_{\rm eq}({\bf k})$, where $c_{\rm eq}({\bf k})$ is the Fourier transform of the equilibrium direct correlation function, in~\eqref{eq:final_expression_gr} to effectively account for higher-order effects due to strong interactions between particles.

\begin{figure*}
    \includegraphics[scale=0.35, clip=True]{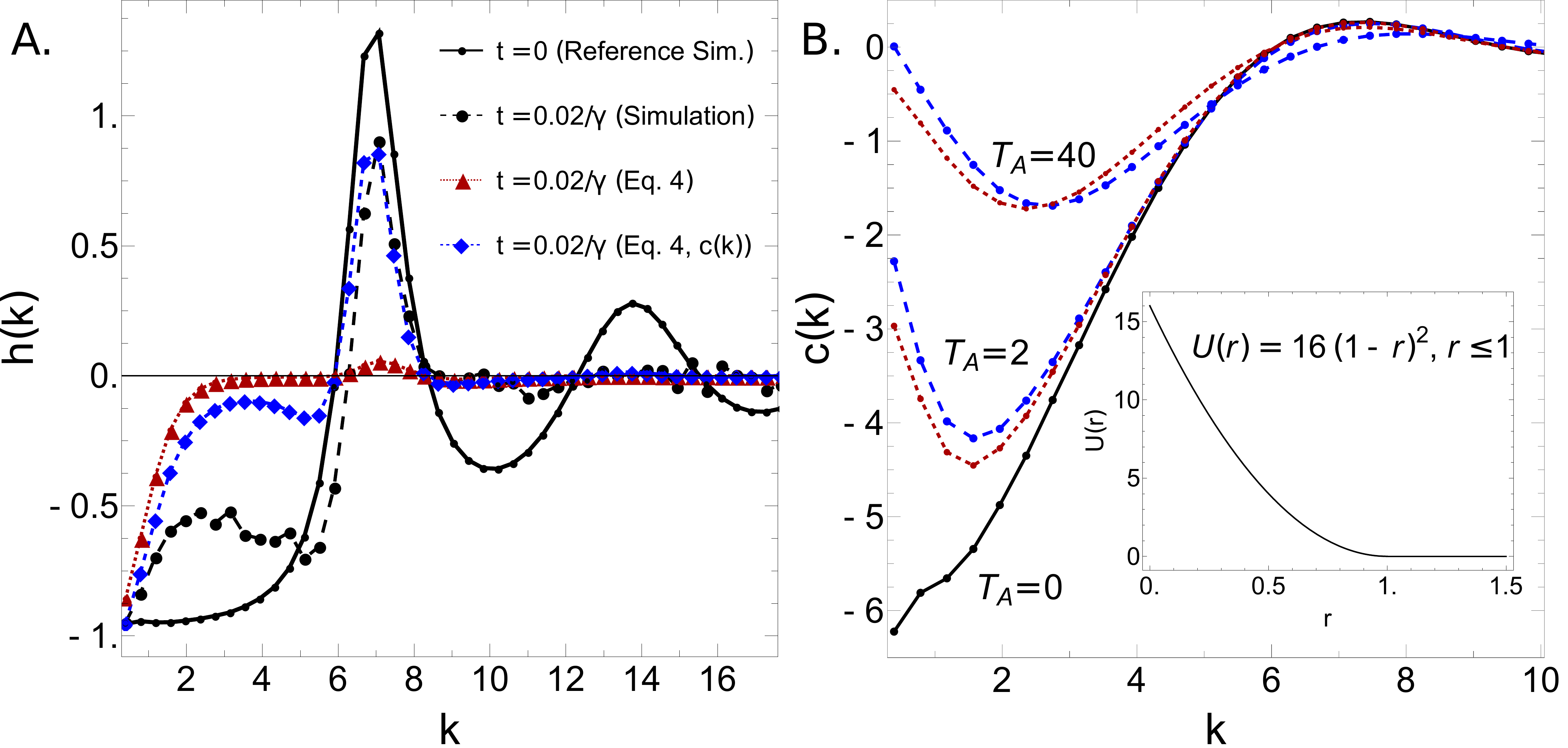}
    \caption{Mean-field theory for nonequilibrium structure for a system of strongly interacting AOUPs (see inset for potential).
	(a)~Relaxation of density correlations $h({\bf k})$ from initial conditions after removing a particle from the system (relaxation time is taken as $0.02/\gamma$). Except at small values of the wavenumber $k=|{\bf k}|$, we find good agreement between measured $h({\bf k})$ (dashed black line with circles) and $h({\bf k})$ predicted from $c_{\rm eq}({\bf k})$ (dashed blue line with diamonds), while predictions using $U({\bf k})$ (dashed red line with triangles) are very poor. Parameters: $\rho_0 = 1.0$, $A = 64$, $\tau = 1.0$, $T_{\rm A}=0$, $T = 1$, $\gamma = 1$.
	(b)~Prediction for the nonequilibrium direct correlation function $c({\bf k})$, as defined in~\eqref{eq:c}. The predicted curves for $c$ (dashed red lines with squares) are compared with simulation results (dashed blue lines with circles) both near equilibrium ($T_{\rm A}=2$, lower pair of dashed lines) and far from equilibrium ($T_{\rm A}=40$, higher pair of dashed lines). The reference $c_{\rm eq}$ (solid black line), which is used as an input for the mean-field prediction, is measured numerically. The good agreement between predictions and simulations demonstrates that our mean-field theory captures well the deviation from equilibrium structure. In particular, it reproduces quantitatively the effective attraction at large wavelengths/small wavenumbers arising due to active forces. Parameters: $\rho_0 = 1.0$, $A = 16$, $\tau = 0.4$, $T = 1$, $\gamma = 1$.
	Note that while the $c_{\rm eq}({\bf r})$-based prediction for the \emph{relaxation of passive particles to a steady state} over short time scales in (a) is poor in the small-k regime, this does not mean that the $c_{\rm eq}({\bf r})$-based theory is poor at predicting the structure of \emph{steady states of active particles} in the same small-k regime in (b). Simulation results for $h({\bf k})$, $c({\bf k})$ are computed from measured $g({\bf r})$ as described in Materials and Methods.}
	\label{Fig:ck}
\end{figure*}

To further motivate applying this approach to our mean-field theory, we investigate the response of a fluid at equilibrium to a time-varying perturbation. Specifically, we simulate a system of particles interacting via the short-ranged repulsive harmonic potential $U({\bf r})=\{A (1- r)^2, r<1; 0, r \geq 1\}$,  with $A=64 T$. We measure the relaxation of density $\rho\left({\bf r} - \tilde{\bf r},t\right)$, where $\tilde{\bf r}$ is the position at which a particle is removed from the system at $t = 0$. We compare this with the predictions obtained both from the linearized dynamics for the density in~\eqref{eq:field_EOM} and from this same equation but with $U({\bf k})$ replaced by $- T c_{\rm eq}({\bf k})$. The results are shown in Fig.~\ref{Fig:ck}(a) for the two-point correlation $h({\bf k})$.

We find that the evolution equation (Eq.~\ref{eq:field_EOM} at $T_{\rm A}=0$) with $U({\bf k})$ substituted by $- T c_{\rm eq}({\bf k})$ yields an accurate prediction for the decay of $h({\bf k})$ in the region of $k$-space around the primary peak of $h_{\rm eq}({\bf k})$. In contrast, predictions obtained \emph{without} using this substitution are very poor in this region, highlighting that systems of strongly interacting particles violate the assumptions underlying linearization of density dynamics. Unsurprisingly, predictions with both methods are poor in the small-$k$ region corresponding to the structure of the system on large length scales. Indeed, changes in structure on such length scales are connected to the compressibility of the active system, which is difficult to predict~\cite{Dulaney2021}. This result numerically shows that the density responses in a strongly interacting fluid are more appropriately captured by the direct correlation function.

Combined with the aforementioned intuition from equilibrium solvation theories, this motivates us to substitute $U({\bf k})$ with $- T c_{\rm eq}({\bf k})$ in~\eqref{eq:field_EOM} and in the subsequent non-equilibrium mean field theory as a heuristic approach to correct for higher-order interactions. Overall, our theory then leads to the following expression for the density correlations:
\begin{align*}\label{eq:final_expression_gr_c}
    h({\bf k}) &= {\bf k}^2 \dfrac{ \hat{G}({\bf k}) + \T}{\hat{G}({\bf k})} T c_{\rm eq}({\bf k}) \int_{-\infty}^0 ds e^{{\bf k}^2 [ (\hat{G}({\bf k}) + \T)s + R(s) ]} ,
    \numberthis
\end{align*}
where $\hat{G}\left({\bf k}\right) =  T\left(1 -\rho_0 c_{\rm eq}\left({\bf k}\right)\right)$ and $R\left(s\right)$ is the same as in~\eqref{eq:expxi2}.

\begin{figure*}
    \includegraphics[scale=0.35, clip=True]{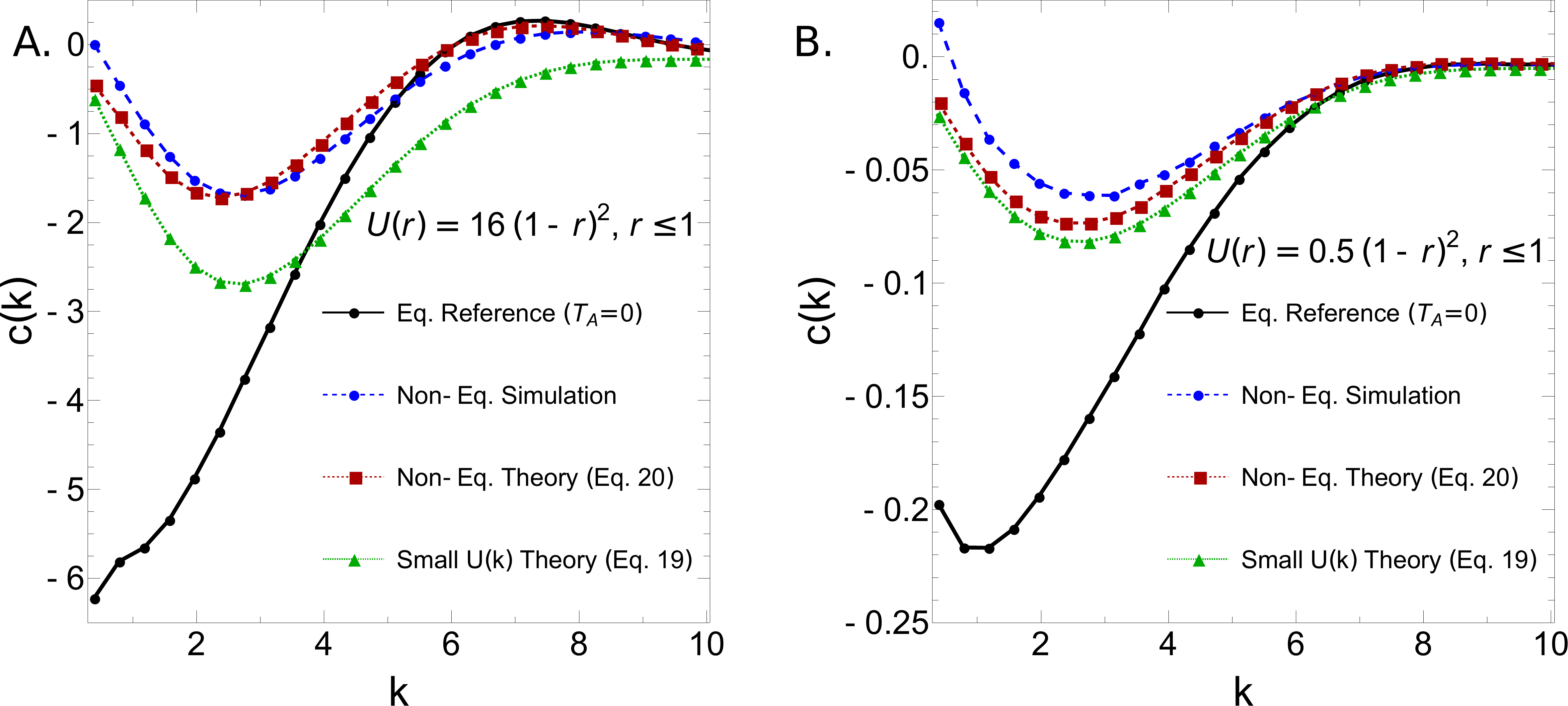}
    \caption{Mean-field theory for nonequilibrium structure for additional systems of strongly interacting AOUPs. 
	(a)~Prediction from mean-field theory for the nonequilibrium direct correlation function $c({\bf k})$, as defined in~\eqref{eq:c}, for AOUPs interacting via harmonic potential with $A = 16$. Details are the same as in Fig.~\ref{Fig:ck}(b), but with only results for $T_{\rm A}=40$ shown, and with an additional estimate for $c({\bf k})$ made using the original mean-field theory~\eqref{eq:final_expression_gr} instead of the version modified by the $c({\bf k})$ substitution~\eqref{eq:final_expression_gr_c} (dashed green line with triangles). The agreement between this new prediction with measurements from simulations is quite poor, as expected for strongly interacting particles to which the original mean-field assumption of a tracer weakly interacting with the bath does not apply and in which nothing is done to effectively account for higher-order interactions.
	(b)~Prediction from both versions of the mean-field theory for the nonequilibrium direct correlation function $c({\bf k})$, as defined in~\eqref{eq:c}, for AOUPs interacting via harmonic potential with $A = 0.5$. All details except potential amplitude are the same as in the first panel. As the AOUPs are interacting much less strongly and the system is closer to the mean-field regime, the prediction from~\eqref{eq:final_expression_gr} is much more accurate, although in fact still less accurate than that obtained via~\eqref{eq:final_expression_gr_c} at all values of ${\bf k}$. This latter prediction is in turn slightly less accurate than the corresponding prediction in the first panel.}
	\label{Fig:ck_2}
\end{figure*}

At equilibrium ($T_{\rm A}=0$),~\eqref{eq:final_expression_gr_c} is equivalent to the famous Ornstein-Zernike relation~\cite{Hansen}. Away from equilibrium ($T_{\rm A}\neq 0$), our prediction~\eqref{eq:final_expression_gr_c} can be used to deduce the structure of the system, given by $h({\bf k})$, based solely on measurements of the equilibrium structure (i.e. from $c_{\rm eq}({\bf k})$). We reiterate that the perturbation theory leading to~\eqref{eq:final_expression_gr_c} ignores the effect of the polarization term in~\eqref{eq:field_EOM}, which was found in Ref.~\cite{Suri2020} to be negligible in a large range of systems and regimes. We surmise that these contributions are small under the set of assumptions, approximations, and regimes that we employ in the present paper as well. Our numerical results support this hypothesis.



To compare our mean-field prediction with numerical results, we introduce the {\it nonequilibrium direct correlation function}, denoted by $c$ and defined as
\begin{equation}\label{eq:c}
	c({\bf k}) = \frac{h({\bf k})}{1+\rho_0 h({\bf k})} .
\end{equation}
This definition can be regarded as a straightforward extension of the Ornstein-Zernike relation for equilibrium liquids, but note that $c$ can no longer be related to any free-energy {\it a priori}. In Fig.~\ref{Fig:ck}(b), we plot the predicted $c$, as deduced from Eqs.~(\ref{eq:final_expression_gr}-\ref{eq:c}), along with measurements obtained from simulations. We again emphasize that the only input for our prediction is the equilibrium direct correlation function $c_{\rm eq}({\bf k})$.

We simulate particles interacting via the short-ranged harmonic potential, given by $U({\bf r})=\{A (1- r)^2, r<1; 0, r \geq 1\}$ with $A=16 T$, at multiple values of $T_{\rm A}$ (Fig.~\ref{Fig:ck}(b)). Our theory accurately predicts the nonequilibrium direct correlation function, particularly in the regime of long wavelengths/small wavenumbers, although there are noticeable discrepancies at higher wavenumbers where the prediction for $c$ deviates insufficiently from $c_{\rm eq}$. To a first approximation, the difference $c-c_{\rm eq}$ can be effectively interpreted as a weak perturbation with respect to the original potential $U$. In other words, $c-c_{\rm eq}$ illustrates how adding active forces to the dynamics affects the microscopic interactions. In the results in Fig.~\ref{Fig:ck}(b), this corresponds to adding an attractive potential, leading to enhanced clustering of particles (and eventually phase separation for particles with sufficiently large driving forces and very strongly repulsive interactions). In Fig.~\ref{Fig:ck_2}, we compare the predictions from the original theory for weakly interacting particles~\eqref{eq:final_expression_gr} (green line, triangles) with predictions from our updated theory~\eqref{eq:final_expression_gr_c}. As shown in the first panel, there is a dramatic improvement in the accuracy of predictions for strongly interacting particles. Note that the non-equilibrium forcing, as parameterized by $T_A=40$, is quite strong, and that our theory is nonetheless able to accurately capture the structure in this regime.

\section{Discussion}

Our results demonstrate that activity-induced changes to the steady-state structure of AOUP systems can be accurately predicted in a wide span of regimes simply from the pair correlations of the system in the absence of activity. Although it is well-known that active forces affect the emerging structure~\cite{Gompper_2020}, reliably predicting the nonequilibrium structure of active systems has remained largely an outstanding problem~\cite{Szamel2015, Rein2016, Wittmann2017b, Szamel2019}. In this work, we propose a mean-field theory which quantitatively predicts the two-point density correlations, illustrating the utility of the direct correlation function in effectively accounting for higher-order interactions.

It would be interesting to explore whether such theories can be extended to other types of active liquids, such as for instance liquids with aligning interactions among the particles~\cite{Chate2020}, or with driving forces that sustain a permanent spinning of particles with isotropic interaction potentials~\cite{Bartolo2016, Vitelli2018}. Since our approach relies mostly on tools of liquid-state theory~\cite{Chandler1993, Hansen, Dean_1996}, which are agnostic to the details of the driving forces, we anticipate that it might be possible to systematically improve our predictions. Thus, we believe that our approach can serve as a basis for developing perturbation theories in generic nonequilibrium liquids~\cite{Nandi2015}. This would open the door to anticipating how density correlations are modified by {\it any} type of driving forces, as a first step towards externally controlling the emerging structure with a specific drive~\cite{England2015, Suri2016, Nagel2020}.

This work was mainly funded by support from a DOE BES Grant DE-SC0019765 to LT and SV (Theory and Machine learning). This research was funded in part by the Luxembourg National Research Fund (FNR), grant reference 14389168.


\section*{Materials and Methods}

\subsection*{Numerical simulations}

Simulations are run in a two-dimensional box $40\sigma \times 40\sigma$ with periodic boundary conditions, where $\sigma = 1$ is the particle diameter. The time step for the simulations is $\delta t = 10^{-4}$. The density was set to $1$ when the harmonic potential is used.

 The equations of motion are integrated using a custom molecular dynamics code based on finite time difference. The systems are equilibrated or allowed to reach a steady state over 500 units of simulation time, corresponding to at least $500 \tau$ for all simulations, where $\tau$ is the persistence time of the active noise, and data is collected every 100 units from the end of equilibration for a duration of 1000 time steps.

Calculation of $c_{\rm eq}({\bf k})$ for theoretical predictions is done by numerically Fourier transforming the portion of the equilibrium $h_{\rm eq}({\bf r}) = g_{\rm eq}({\bf r}) - 1$ with $r \in [0, 16]$ to obtain $h_{\rm eq}({\bf k})$, and then computing $c_{\rm eq}({\bf k})$ using the Ornstein-Zernike relation (shown in~\eqref{eq:c} extended to non-equilibrium systems). Equilibrium and non-equilibrium $g({\bf r})$ is obtained by generating histograms of distances between each pair of particles with resolution $dr = 0.01\sigma$, averaged over 15 independent trials with 11 snapshots per trial and limited to $r \in [0, 20]$. Fourier transformation to obtain $h({\bf k})$ is done by multiplying $h({\bf r})$ by $2 \pi r J_0(k \cdot r)$ and integrating over $r \in [0, 16]$, repeated for $k \in [2\pi/16, 16\pi]$ incremented by $2\pi/16$.

Perturbation simulations to obtain data in Fig.~\ref{Fig:ck}(a) at $t \neq 0$ are equilibrated for 99.8 units of simulation time, then measured every 0.02 units of time for an additional 0.2 units, as this was found to include all of the measurable relaxation behavior. For each `snapshot' separated by 0.02 units of time, $g\left({\bf r} - \tilde{\bf r}\right)$ is obtained by generating histograms of the distance of each particle from $\tilde{\bf r}$ with resolution $dr = 0.01\sigma$, averaged across 10 consecutive time steps for each snapshot and over at least 500 independent trials.

\subsection*{Code}

Codes for molecular dynamics can be found at \url{https://github.com/ltociu/structure_dissipation_active_matter}.

\bibliography{References}
\end{document}